\documentclass[12pt]{article}
\usepackage{epsfig}

\newcommand\pubnumber{SLAC-PUB-9616}
\newcommand\pubdate{February, 2003}
\newcommand\hepnumber{physics/0302044}

\def\SLAC{Stanford Linear Accelerator Center\\
    Stanford University, Stanford, California 94309 USA}
\def\doeack{\footnote{Work supported by the Department of Energy,
                     contract DE--AC03--76SF00515.}}

\def\Title#1{\begin{center} {\Large #1 } \end{center}}
\def\Author#1{\begin{center}{ \sc #1} \end{center}}
\def\Address#1{\begin{center}{ \it #1} \end{center}}

\def\submit#1{\begin{center}Submitted to {\sl #1} \end{center}}
\newcommand\pubblock{\rightline{\begin{tabular}{l} \pubnumber\\
         \pubdate \\ \hepnumber \end{tabular}}}
\newenvironment{Abstract}{\begin{quotation} \begin{center}
                       ABSTRACT
     \end{center}\bigskip  }{\end{quotation}}
\def\submit#1{\begin{center}Submitted to {\sl #1} \end{center}}
\def\Acknowledgements{\bigskip  \bigskip \begin{center} \begin{large}
             \bf ACKNOWLEDGEMENTS \end{large}\end{center}}

\def\beq{\begin{equation}}
\def\eeq#1{\label{#1}\end{equation}}
\def\eeqn{\end{equation}}

\newenvironment{Eqnarray}%
   {\arraycolsep 0.14em\begin{eqnarray}}{\end{eqnarray}}
\def\beqa{\begin{Eqnarray}}
\def\eeqa#1{\label{#1}\end{Eqnarray}}
\def\eeqan{\end{Eqnarray}}
\def\CR{\nonumber \\ }

\let\bar=\overbar

\def\eg{{\it e.g.}}

\def\lsim{\mathrel{\raise.3ex\hbox{$<$\kern-.75em\lower1ex\hbox{$\sim$}}}}
\def\gsim{\mathrel{\raise.3ex\hbox{$>$\kern-.75em\lower1ex\hbox{$\sim$}}}}

\def\half{\frac{1}{2}}

\def\del{\partial}
\def\Dslash{\not{\hbox{\kern-4pt $D$}}}
\def\dslash{\not{\hbox{\kern-2pt $\del$}}}

\def\msb{{\bar{\ssstyle M \kern -1pt S}}}

\makeatletter
\def\section{\@startsection{section}{0}{\z@}{5.5ex plus .5ex minus
 1.5ex}{2.3ex plus .2ex}{\large\bf}}
\def\subsection{\@startsection{subsection}{1}{\z@}{3.5ex plus .5ex minus
 1.5ex}{1.3ex plus .2ex}{\normalsize\bf}}
\def\subsubsection{\@startsection{subsubsection}{2}{\z@}{-3.5ex plus
-1ex minus  -.2ex}{2.3ex plus .2ex}{\normalsize\sl}}

\renewcommand{\@makecaption}[2]{%
   \vskip 10pt
   \setbox\@tempboxa\hbox{\small #1: #2}
   \ifdim \wd\@tempboxa >\hsize     
       \small #1: #2\par          
     \else                        
       \hbox to\hsize{\hfil\box\@tempboxa\hfil}
   \fi}

 \def\citenum#1{{\def\@cite##1##2{##1}\cite{#1}}}
 
\newcount\@tempcntc
\def\@citex[#1]#2{\if@filesw\immediate\write\@auxout{\string\citation{#2}}\fi
  \@tempcnta\z@\@tempcntb\m@ne\def\@citea{}\@cite{\@for\@citeb:=#2\do
    {\@ifundefined
       {b@\@citeb}{\@citeo\@tempcntb\m@ne\@citea\def\@citea{,}{\bf ?}\@warning
       {Citation `\@citeb' on page \thepage \space undefined}}%
    {\setbox\z@\hbox{\global\@tempcntc0\csname b@\@citeb\endcsname\relax}%
     \ifnum\@tempcntc=\z@ \@citeo\@tempcntb\m@ne
       \@citea\def\@citea{,}\hbox{\csname b@\@citeb\endcsname}%
     \else
      \advance\@tempcntb\@ne
      \ifnum\@tempcntb=\@tempcntc
      \else\advance\@tempcntb\m@ne\@citeo
      \@tempcnta\@tempcntc\@tempcntb\@tempcntc\fi\fi}}\@citeo}{#1}}
\def\@citeo{\ifnum\@tempcnta>\@tempcntb\else\@citea\def\@citea{,}%
  \ifnum\@tempcnta=\@tempcntb\the\@tempcnta\else
  {\advance\@tempcnta\@ne\ifnum\@tempcnta=\@tempcntb \else\def\@citea{--}\fi
    \advance\@tempcnta\m@ne\the\@tempcnta\@citea\the\@tempcntb}\fi\fi}
\makeatother

\def\t#1{{\tt #1}}

\begin{document}
\begin{titlepage}
\pubblock
\vfill
\Title{Abstract Applets:\\
        a Method for Integrating Numerical Problem-Solving\\
       into the Undergraduate Physics Curriculum}
\vfill
\Author{Michael E. Peskin\doeack}
\Address{\SLAC}
\vfill
\begin{Abstract}
In upper-division undergraduate physics courses, it is desirable to give
numerical problem-solving exercises integrated naturally into weekly problem 
sets.  I explain a method for doing this that makes use of the built-in class 
structure of the Java programming language.  I also supply a 
Java class library that
can assist instructors in writing programs of this type.
\end{Abstract}
\vfill
\submit{Computing in Science and Engineering}
\vfill
\end{titlepage}
\tableofcontents
\newpage
\def\thefootnote{\fnsymbol{footnote}}
\setcounter{footnote}{0}

\section{Introduction}

A typical upper-division undergraduate physics course, for example, the 
course in classical electrodynamics at the level of Griffiths' 
text~\cite{Griffiths},
involves extensive analytic problem solving.  Students learn to solve Laplace's
equation in rectangular coordinates, in cylindrical coordinates, with Bessel 
functions, with Legendre polynomials.  The methods taught are important 
in their
own right and as methods to illustrate central physical concepts.  But it would
be better if numerical problem-solving methods were discussed at a similar
level.  Students today typically have in their dorm rooms, and even carry
 in their 
backpacks,
computers with the power of a typical university central mainframe of the 
1970's.
It would be wonderful to put this power to work
both to developing intuition and in explicit problem-solving.

The inclusion of numerical calculations in the physics curriculum is even more 
important because our students typically go to careers, either in research
or in industrial settings, in which the basic task is modelling of physical 
systems.
Analytic methods are useful for estimates, or for working out the dependence
on parameters, but understanding a realistic system in detail typically 
requires a 
computer simulation.  So we should make clear to our
students right away that it is 
straightforward
 to put the equations that appear in their classes
 onto  a computer and obtain sensible physical
results.

This problem is currently addressed in the physics curriculum in two ways.  
First,
specific physics problems are encoded in computer programs which 
students can run as black boxes, changing the parameters and seeing 
what consequences develop.  The `scripted applets' of the Davidson College
group represent a very beautiful development along this line~\cite{scripted}. 
 Such black-box
programs are useful in introductory courses, but, for upper-division 
courses,
they do not teach all of the skills one would like to develop.  These 
programs are
time-consuming to write, and it is often the philosophy that students 
should not
touch the code.  (The Davidson
 applets are designed so that not even professors 
modify the code.)  But we would like students to write some code, and to 
understand how simple sets of instructions can iterate to the patterns 
that solve
interesting equations of Nature.   Often in undergraduate mechanics courses,
instructors give students differential equations to be integrated by
commercial packages such as Maple or 
Mathematica.   Such projects have a similar black-box approach.

The other common approach is a computational physics course, on the model of
the textbooks of Gould and Tobochnik~\cite{GT} and Koonin~\cite{koonin}.
  Such courses are often 
constructed around major projects.  Students spend a large part of a semester
learning a learing a computer environment, then construct an elaborate code
for one particular application.  Such a course is important to give students
experience with large-scale computer applications and to begin studying 
sophisticated numerical methods. But it  would be good 
also to allow students to 
do simpler numerical problems that tie in directly to 
material being covered in 
their core courses. 

 Ideally, a weekly problem set in mechanics or electrodynamics
should consist of several analytical and one numerical problem.  The
reason this is not done as a matter of course is the  difficulty of 
having students
master the purely computer programming
aspects of the task.

In this article, I propose a way of solving this problem.  
The method is based on the
structure of the programming language Java.  
Java is an object-oriented computer language which allows programs that 
have a hierarchial structure.   Thus, 
one can assign to students the writing of a small piece of code, perhaps one
subroutine that carries out a numerical computation.  This code can then fit 
together with a larger program that implements a graphical user interface for
visualizing the results of the program.  The
user interface  code can be assembled by the 
instructor; students need not be bothered by its complexity.  Finally, the 
instructor's task in writing this larger program can be simplified 
if the various
user interface elements are drawn from a pre-assembled `class library'. 

At heart, this is the familiar strategy of asking students to write a 
subroutine
which can be tied to a larger package of code.  The use of Java assists this 
in two ways.  First, it allows the instructor to neatly encapsulate the 
part of the
code for which the student is responsible, hiding the details of the graphical 
elements and user interface.  Second, it makes available to the instructor a
programming library that makes it rather simple to write graphical elements
that are interesting and pleasing. 

 This 
strategy could also be carried out in other programming languages that allow
easy access to graphical user interface components, for example, Visual Basic.
The CUPS project has created a range of pleasing simulations in 
Pascal~\cite{CUPS}, and these are extensible by working at the code level.
Java has the advantage that the same code runs on UNIX, Windows, or 
Macintosh systems, so that students can put together their 
assignments on whatever operating system is most convenient for them.
In addition, Java belongs to a family of computer languages, including 
\t{C}, \t{C++}, and Pascal, in which numerical computations have a common 
syntax.  The part of the code that the student must write is almost 
indistinguishable
among these four languages, so their is no need for students to have 
prior experience in Java.

An extensive set of software resources for creating educational simulations
in Java is being put together by Christian, Gould, and Tobochnik~\cite{CGT}
as a part of their work for the second edition of the textbook~\cite{GT}.
In contrast to that work, I provide here a minimal set of Java resources
to provide exercises of the type I have described.

In this paper, I present this method of constructing problems 
with numerical
solutions.  In Section 2, I present a sample numerical exercise 
that I have 
given on a problem set.  In Section 3, I give some further 
examples of such
numerical exercises.  All of these examples are based on a relatively
simple class library of graphical elements that are specifically useful 
for programming exercises of this type.   In Section 4, I describe the 
basic structure and format of this library.  To illustrate its 
features, the master program for the first exercise is discussed
in detail.   Section 5 presents some conclusions.
Appendix A gives the complete documentation for the class library.
Appendix B gives a list of the example programs included in the 
software distribution.

The Java code for the class library and for the example programs discussed
in this article can be found in a \t{tar} file at the web site at which 
the eprint of this paper is posted~\cite{eprint}.

\section{Laplace Applet}

Textbook discussions of Laplace's equation (e.g. \cite{Griffiths}) note the
fact that a solution $\phi(x)$ is the average of the solution at 
neighboring points, 
and that this fact can be the basis for a numerical solution of 
Laplace's equation.
In this section, I will present a homework problem that allows a student to 
implement this observation in a numerical program and see it work.

Look at the computer program shown in Fig. \ref{fig:LaplaceApp}.  It is not so 
difficult to make sense of this program.  Its idea is to sweep through an array
\t{phi[i][j]}, updating successive values.  The programs tests for the maximum 
change over the array and quits when this is sufficiently small.  The algorithm
for updating the array is not given.  But one can explain in words that each 
array element can be, successively, set equal to the average of its neighbors, 
and that the equilibration of this process yields a 
solution of the Laplace equation.

To implement this algorithm, it is necessary to 
modify only one line of the code,
replacing the assignment to \t{newphi} by 
\beqa
 & & \verb!double newphi = ! \CR
 & & \hskip 0.3in 
\verb!(0.25)*(phi[i+1][j]+phi[i][j+1]+phi[i-1][j]+phi[i][j-1]);! \ .
\eeqa{newphi}
This statement has the same form in  \t{C}, \t{C++}, Pascal, or Java.  
It should 
be at least recognizable by students whose only programming experience is 
in Basic or FORTRAN.

\begin{figure}[p]
\begin{center}
\begin{verbatim}
import java.awt.*;
import java.awt.event.*;
import java.applet.Applet;

public class Laplace extends LaplaceGUI{
  double criterion = 1.0e-2;

  void solve(){
    double maxdiff = 1.0;
    int iteration = 0;
    while ( maxdiff > criterion){
      for (int n = 1; n <= 20; n++){
        maxdiff = 0.0;
        for (int i = 1; i < Nx; i++){
          for (int j =1 ; j < Ny; j++){
              /* check whether (i,j) is a cathode or ground point */
            if (normal(i,j) == false) continue;
              /* update the phi array  */
            double oldphi = phi[i][j];
              /*  put something more sensible here  : */
            double newphi = 33.0;
            phi[i][j] = newphi;
	    /*  compute the criterion for stopping */
            double delta = Math.abs(newphi-oldphi);
            if (delta > maxdiff) maxdiff = delta;
          }
        }
        iteration++;
      } 
      refreshPicture();
      Legend.write("max. diff : "+maxdiff+ "      "+iteration);
      if (timetostop) break;
    }
  }
\end{verbatim}
\caption{The program \t{Laplace.java}.}
\label{fig:LaplaceApp}
\end{center}
\end{figure}

The statement \t{class Laplace extends LaplaceGUI} indicates that the simple
program \t{Laplace.java} 
shown in Fig. \ref{fig:LaplaceApp} is intended to work with functions and
data structures defined in a parent program \t{LaplaceGUI.java}.  
This program, its parent \t{PhysicsApplet.java}, and a small 
file \t{Laplace.html} are given the directory \t{Laplace} in the 
software distribution mentioned at the end of the Introduction. These
files
can be made available
 for download on the course Web page.  Compiling these programs
together, one obtains a working simulation toy.  The method of 
linking the programs depends on the precise operating system, but 
under UNIX (or Mac OS 
X) it is as simple as putting the four files in the same directory and typing
\t{javac Laplace.java}.  The compiled program, or `applet', is then run by 
viewing the file \t{Laplace.html} with  a  Web browser.   The student would not
be expected to modify, or even open, any of these files except for the original
\t{Laplace.java}.   This file contains all of the physics; the others simply 
supply the computer interface.

\begin{figure}[tb]
\begin{center}
\epsfig{file=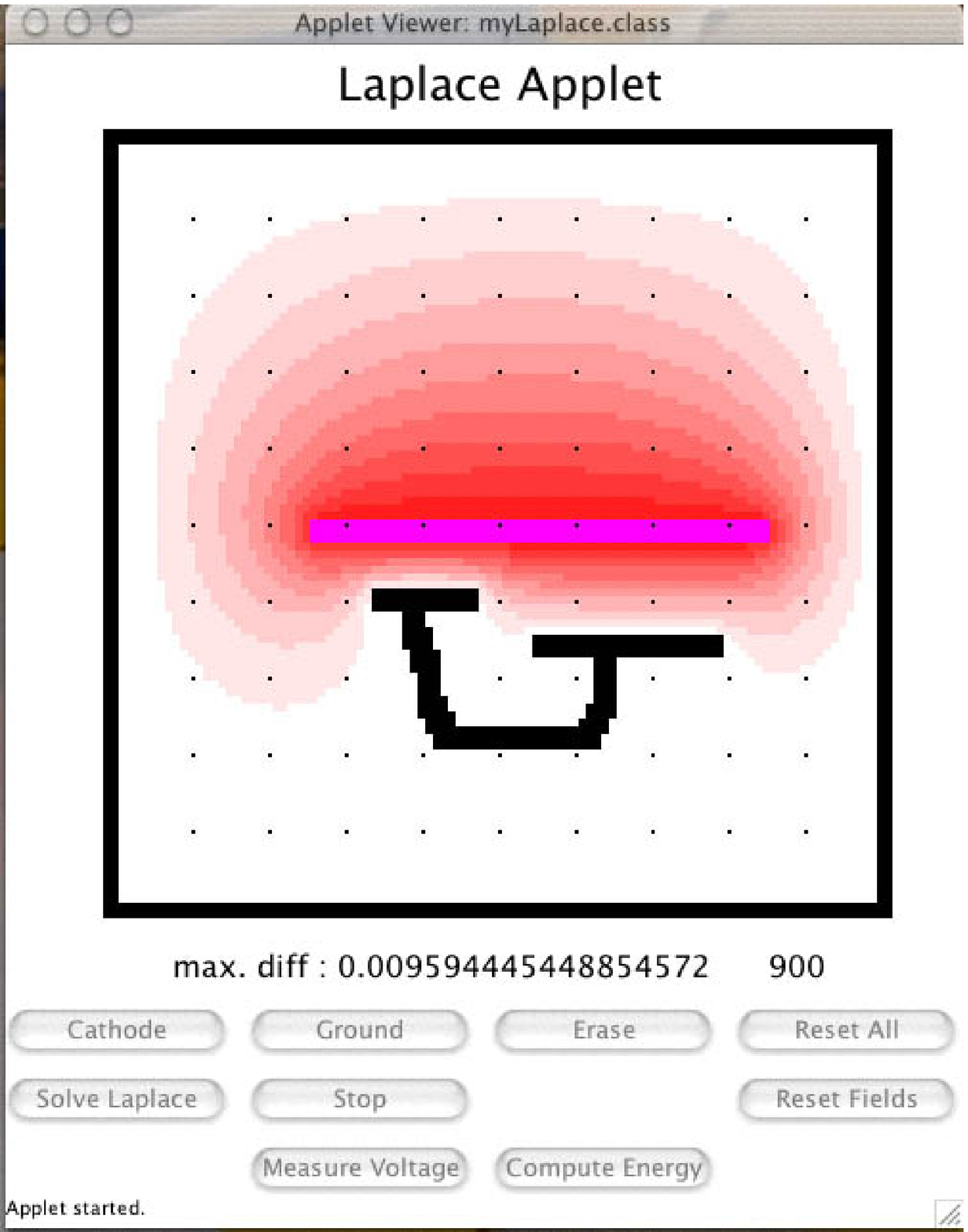,height=5.0in}
\caption{Working version of the \t{Laplace} applet.}
\label{fig:Laplace}
\end{center}
\end{figure}

The result of this process is a working application, a view of which is 
shown in Fig.~\ref{fig:Laplace}.
 The applet shows a figure with a box that displays the values of the
array \t{phi} in greyscale.  In the specific example shown, 
\t{phi} is chosen to 
be a $100\times 100$ array.   The box contain dots as fiducial marks
at
each tenth grid point, to facilitate specific numerical computations.
By clicking on the buttons `Cathode' and `Ground', one
can paint a set of boundary conditions with the mouse.  Clicking on the button
`Solve Laplace' calls the method \t{solve()} in the program \t{Laplace.java}.
As the array \t{phi[i][j]} is updated, the values of 
\t{phi} are displayed on the 
screen in grayscale.  Clicking on the button 
`Measure Voltage' and then clicking
on the screen causes the value of \t{phi} at that point to appear as a label
under the box.

Once the applet is programmed and working correctly, it can be used for 
many illuminating exercises.  Some of these are  qualitative problems, for 
example, illustrating the principle of a Faraday cage by placing small 
grounded conductors
around a cathode.  Others are quantitative, for example, working out the size
of edge effects on the capacitance of a capacitor of finite size.  
To aid in the
latter calculation, the button `Compute Energy' computes the electrostatic 
energy stored in the configuration shown from a discrete approximation to 
the expression $\int d^2x \half \epsilon_0 E^2$.   The problem set that 
contained
this applet asked the student to implement a function \t{Energy()} 
that would be
called by this button and return the result.  The GUI would then take care of 
writing this result to the screen.

\section{Further Examples}

Many other computational exercises can be constructed along these lines.
Eight additional applets are included with the software distribution.
I describe two of these below.

\begin{figure}[tb]
\begin{center}
\epsfig{file=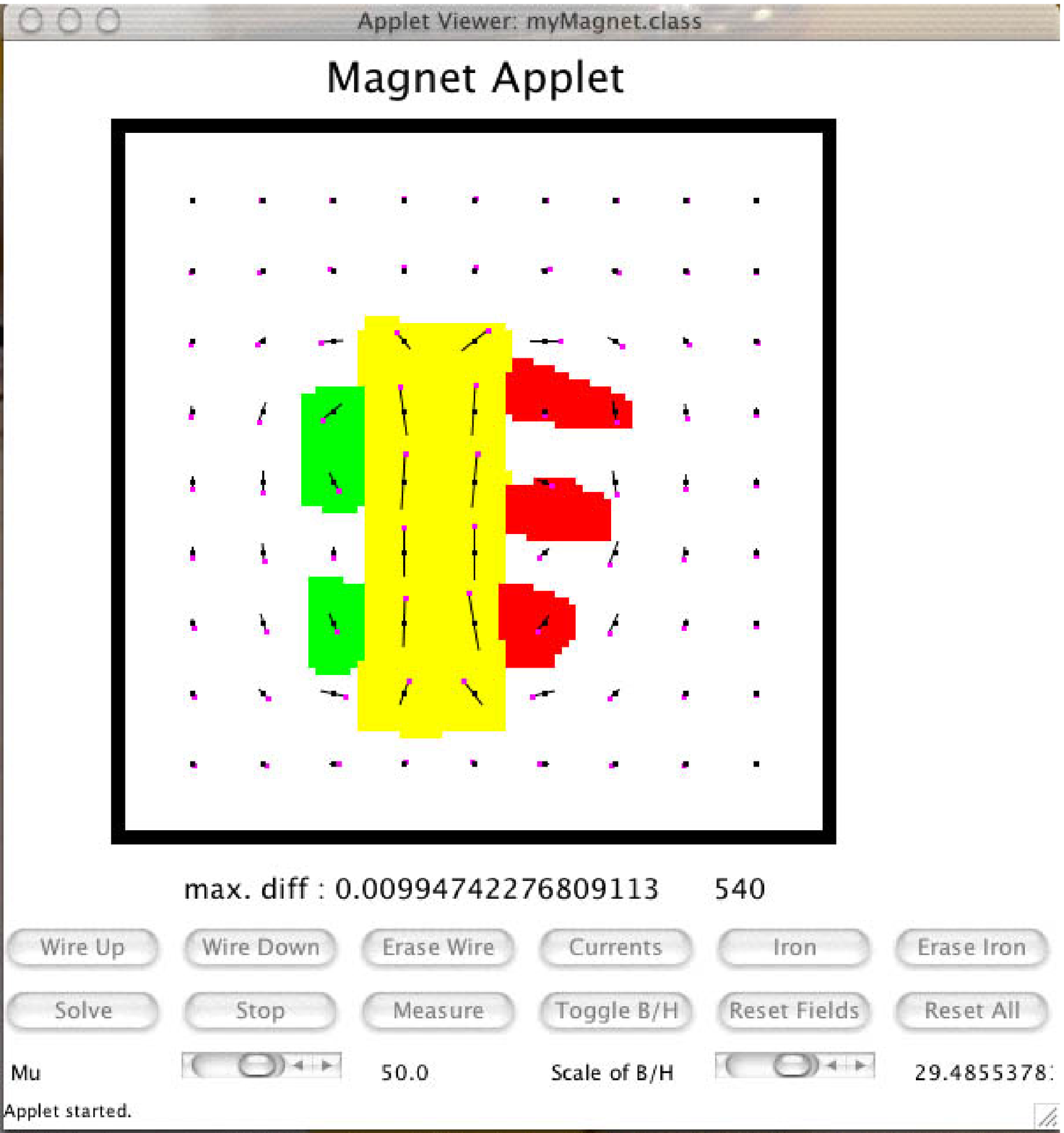,height=2.8in}\hskip 0.2in
\epsfig{file=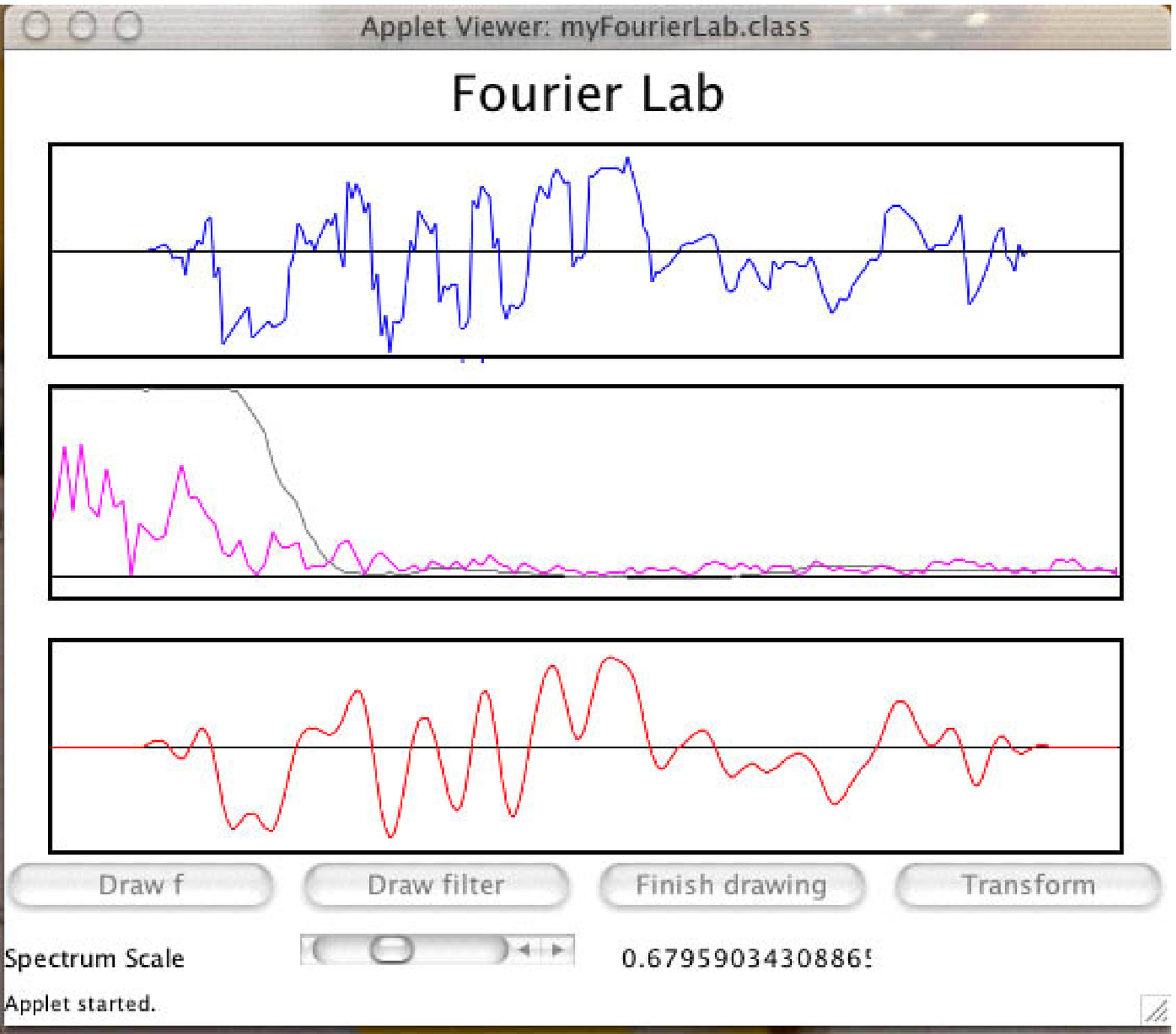,height=2.5in}
\caption{Working versions of the \t{Magnet}  (left) and \t{FourierLab} 
(right) applets.}
\label{fig:Magnet}
\end{center}
\end{figure}

The left-hand side of Fig.~\ref{fig:Magnet} shows an applet 
that computs the magnetic 
fields in an array of
wires and magnetic material that can be drawn on the screen.  The situation is
uniform in  the third dimension, with current flowing up or 
down through the 
screen.  A fixed current, up or down, is assigned to squares colored,
respective, green or red.  These squares can be painted with the 
mouse.
The student can also color in regions to be filled with a
linear magnetic material (`iron') with
adjustible permeability $\mu$.  The magnetic fields are generated from a
vector potential $   \vec A = ( 0 , 0,  {\cal A})$, where ${\cal A}$ 
solves the 
Poisson equation
\beq
              - \nabla^2 {\cal A} =  \mu {\cal J} \ ,
\eeqn
in which  ${\cal J}$ is the current density in the wires (in A/m$^2$).  
The equation
can be solved by the same relaxation method discussed in the previous section
for the Laplace equation.  The problem set containing this applet explained
the strategy, then asked the students to work out the details using their
experience with the \t{Laplace} applet of Section 2.  They were then asked to
use this \t{Magnet} applet to to solve qualitative
examples in the theory of magnetism and quantitative problems of magnet 
design.

The right-hand side of Fig.~\ref{fig:Magnet} shows an applet 
that can be used to illustrate the basic 
principles of signal analysis.  The applet displays three boxes on the
screen. In the upper box, a waveform can be entered,
either from the computer program as a mathematical function or by drawing 
with the mouse.  The center box shows the (modulus of the)
Fourier transform of the waveform.  This is multiplied by a 
filter function which,
again, can either be supplied by the program or drawn on the screen.
To present this applet on a problem set, I supplied to the students five 
files: \t{FourierLab.java},
 \t{FourierTransform.java}, \t{FourierLabGUI.java},
\t{PhysicsApplet.java}, and \t{FourierLab.html}.  The last of 
these is simply used
to display the compiled program.  The file \t{FourierLab.java} contains 
the functions defining the initial waveform and filter. The file 
\t{FourierTransform.java} contains the methods called by the button 
`Transform'.  The remaining two Java programs define the graphical user
interface and are not meant to be modified by the student.  They do not contain
any of the physics of the computation.

The contents of \t{FourierTransform.java} given to the student are shown in 
Fig.~\ref{fig:FTprogram}.  The structure is based on notions of object-oriented
programming, but it is self-explanatory.  The real-space function 
is a real-valued function $f(x)$, represented as an array \t{f[n]}, 
$n = 0, \ldots,
(N-1)$. The 
real and imaginary parts of the Fourier Transform are represented as 
arrays \t{fcos[m]} and \t{fsin[m]}, $m = 0, \ldots, N/2$. Using only the 
basic understanding of the treatment of arrays in any familiar programming
language, it is straightforward to complete the Fourier transform algorithms
so that they work correctly.  (Java does have the peculiar feature that 
$\sin x$, $\cos x$ and $\pi$ are written \t{Math.Sin(x)}, \t{Math.Cos(x)},
and \t{Math.PI}; students must be told about this.)
In principle, one could build different versions of \t{FourierTransform.java}
from this basic structure,
incoporating different algorithms for computing the Fourier Transform.

\begin{figure}[p]
\begin{center}
\begin{verbatim}
public class FourierTransform{

    int Nx, N2; 
    double[] f, fcos, fsin;

    FourierTransform(int N){
        Nx = N;
        N2 = Nx/2;
        f = new double[Nx];
        fcos = new double[N2+1];
        fsin = new double[N2+1];
    }

    void Transform(){
        for (int i = 0; i <= N2 ; i++){
	    double Tc = 0.0;
	    double Ts = 0.0;
            for (int j = 0; j < Nx ; j++){
		/*  do something intelligent here   */
            }
            fcos[i] = Tc;
            fsin[i] = Ts;
	}
    }

    void InverseTransform(){
        for (int i = 0; i < Nx ; i++){
	    double T = fcos[0] + fcos[N2];
            for (int j = 1; j < N2 ; j++){
		/*  do something intelligent here   */
            }
            f[i] = 0.0;
	}
    }
}
\end{verbatim}
\caption{The class {\tt FourierTransform.java}, as provided to the student.}
\label{fig:FTprogram}
\end{center}
\end{figure}

\section{Construction of the Laplace Applet}

The computer programs described in the previous two sections are 
constructed from a simple program, containing the actual physics and 
accessible to the
student, and a more complex user interface program whose contents the student
can ignore.  Up to now, I have not discussed any aspect of the Java 
programming language.  However, the object-oriented nature of this language
is the key to the way that these programs are structured.  
In this section, I will
first review some notions of object-oriented programming and then explain 
how these ideas are applied in the specific case of the Laplace applet.

In object-oriented programming, the basic element of a program is a `class',
a set of data variables together with function (`methods') that act on these
variables.  One class can be created from another in a parent-daughter
relation.  The daughter class is said to `inherit' the variables and methods of
the parent; new variables and methods intrinsic to the daughter can be added
to these.  In this way, a complex structure can be built up in stages.  
It is possible to define an `abstract class' in which one or more methods
are defined in principle but are not implemented.  An abstract class cannot be
created (`instantiated') in a computer program.  However, 
a daughter of the abstract class
which defines the required methods can be created and used.   In Java,
each individual class \t{A} is defined in a separate file \t{A.java}.

The Java language defines an `applet', a mini-program accessible through a 
Web browser, as a predefined parent class.  This \t{Applet} class manages the 
window in which the program appears and provides methods to draw in this
window. 

The definition of the Laplace applet starts from a parent 
class \t{PhysicsApplet}
which is a daughter class of \t{Applet}.
This class defines various user interface elements that are
 useful in constructing
problem set applets of the type that I have presented above.  A complete
documentation of the classes, variables, and methods of \t{PhysicsApplet} is
given in Appendix A.  \t{PhysicsApplet} is an abstract class that leaves a 
large number of methods undefined.  The Java language contains many hooks
to graphical elements, making it straightforward to construct
 basic user interfaces.
There are many excellent books that describe user interface programming
in Java~\cite{HandC,Flanagan}  However, I hope that the collection
of specialized elements contained in \t{PhysicsApplet} will 
make it even a step 
easier for instructors to build their own programs of the 
type illustrated above.

The class \t{LaplaceGUI} is constructed as a daughter class of 
\t{PhysicsApplet}.
The construction this class puts into the applet the specific displays and 
buttons described in Section 2. 
\t{LaplaceGUI} is still an abstract
class.  It defines almost all of the methods of \t{PhysicsApplet} but leaves
undefined the methods \t{solve()}, which actually carries out the solution of 
Laplace's equation, and \t{Energy()}, which computes the electrostatic
energy.  When the class \t{Laplace} adds a definition of 
\t{solve()} to this structure, as shown in Fig.~\ref{fig:LaplaceApp}, 
and also a definition of \t{Energy()}, all needed
methods are defined and the class can be instantiated.  In the writing of 
the \t{Laplace} class, the definitions of all of the other 
methods of \t{LaplaceGUI}
can remain hidden.

The code for the class \t{LaplaceGUI} is shown in Figs 5--8.  The program
is lengthy, but it is all straightforward bookkeeping of the graphical
elements.

The program begins by defining the basic array \t{phi} involved in the 
numerical computation, an array of integers \t{State} that keeps track of the
boundary conditions drawn by the user, a graphic element of type
\t{ArrayDisplay}, which produces the large square in described for the
applet.
and a graphic element of type \t{TitleBanner}, which displays a caption under
this square.

The next segment of the program defines integer constants
that label the options for the behavior of various elements of the program.
Books on graphical user interfaces (notably \cite{Interface}) ask that a user
interface be `modeless', so that the user has as many options as possible at
any given time. The price of this feature is a complex programming style.
My philosophy is just the reverse.  I would like to make the programming task
as easy as possible for the instructor, even if this 
costs the user some flexibility.
Thus, the programming style is completely modal.

\begin{figure}[p]
\begin{center}
\begin{verbatim}
import java.awt.*;
import java.awt.event.*;
import java.applet.Applet;

abstract public class LaplaceGUI extends PhysicsApplet{

  double[][] phi;
  int[][] State;
  ArrayDisplay D;
  TitleBanner Legend;
  double CathodeV = 100.0;
     // mouse modes:
  static final int NormalMode = 0;
  static final int CathodeMode = 1;
  static final int GroundMode = 2;
  static final int EraseMode = 3;
  static final int MeasureMode = 4;
     // button codes:
  static final int FieldResetCode = 1;
  static final int AllResetCode = 2;
  static final int MeasureECode = 3;
  static final int StartCode = 4;
  static final int StopCode = 5;
    // array states:
  static final int NormalState = 0;
  static final int CathodeState = 1;
  static final int GroundState = 2;
  
  LaplaceGUI(){
    super("Laplace Applet",100,100,4,100);
    phi = new double[Nx+1][Ny+1];
    State = new int[Nx+1][Ny+1];
    for (int i = 0; i <= Nx; i++){
      for (int j = 0; j <= Ny; j++){
         phi[i][j] = 0.0;
         State[i][j] = 0;
       }
    }
  }
\end{verbatim}
\caption{The class {\tt LaplaceGUI.java}, part 1}
\label{fig:PAprogram}
\end{center}
\end{figure}

\begin{figure}[p]
\begin{center}
\begin{verbatim}
  void resetArrays(){
      for (int mx = 0; mx <= Nx; mx++){
        for (int my = 0; my <= Ny; my++){
          if (State[mx][my] == 0)  phi[mx][my] = 0.0;
        }
      }
      refreshPicture();
    }

    void resetAll(){
      for (int mx = 0; mx <= Nx; mx++){
        for (int my = 0; my <= Ny; my++){
         phi[mx][my] = 0.0;
         State[mx][my] = 0;
        }
      }
      refreshPicture();
    }

  void buildPicture(){
    D = new ArrayDisplay(phi,State, Color.red,Color.black);
    Picture.add(D,"Center");
    Legend = new TitleBanner(" ", 16);
    Picture.add(Legend,"South");
  }

  void refreshPicture(){
    D.refresh();
  }
\end{verbatim}
\caption{The class {\tt LaplaceGUI.java}, part 2}
\label{fig:PAprogramtwo}
\end{center}
\end{figure}

\begin{figure}[p]
\begin{center}
\begin{verbatim}
  void buildControls(){
    Controls.setLayout(new GridLayout(0,4,10,10));
    ModeButton B1 = new ModeButton("Cathode", CathodeMode);
    Controls.add(B1);
    ModeButton B2 = new ModeButton("Ground", GroundMode);
    Controls.add(B2);   
    ModeButton B3 = new ModeButton("Erase", EraseMode);
    Controls.add(B3);   
    CommandButton B4 = new CommandButton("Reset All", AllResetCode);
    Controls.add(B4);
    CommandButton B5 = new CommandButton("Solve Laplace", StartCode);
    Controls.add(B5);
   CommandButton B6 = new CommandButton("Stop",StopCode);
    Controls.add(B6);
    Label B7 = new Label(" ");
    Controls.add(B7);
    CommandButton B8 =
                 new CommandButton("Reset Fields",FieldResetCode);
    Controls.add(B8);
    Label B9 = new Label(" ");
    Controls.add(B9);
    ModeButton B10 = new ModeButton("Measure Voltage", MeasureMode);
    Controls.add(B10);
    CommandButton B11 =
                 new CommandButton("Compute Energy", MeasureECode);
    Controls.add(B11);
    Label B12 = new Label(" ");
    Controls.add(B12);
  }

  Color findColor(double A, int S){
    if (S == CathodeState)  return D.Color2;
    if (S == GroundState)   return D.Color1;
    int colorm =  (int) ( 10.0 * A/CathodeV);
    return D.DisplayColors[colorm];
  }

  void writeFieldValue(double Val, int mode){
    if (mode == MeasureMode) Legend.write(" Voltage = " + Val);
  }
\end{verbatim}
\caption{The class {\tt LaplaceGUI.java}, part 3.}
\label{fig:PAprogramthree}
\end{center}
\end{figure}

\begin{figure}[p]
\begin{center}
\begin{verbatim}
  void setArrays(int i, int j, int mode){
    if (mode == CathodeMode){
      phi[i][j] = CathodeV;
      State[i][j] = CathodeState;
    } else if (mode == GroundMode){
      phi[i][j] = 0.0;
      State[i][j] = GroundState;
    } else if (mode == EraseMode){
      phi[i][j] = 0.0;
      State[i][j] = 0;
    }
  }

  void doAction(int Code){
    switch(Code){
      case FieldResetCode:
        resetArrays();
        break;;
     case AllResetCode:
        resetAll();
        break;
     case MeasureECode:
        double E = Energy();
        Legend.write("Energy = "+E);
        break;
     case StartCode:
        startThread();
        break;
     case StopCode:
        stopThread();
        break;
     default:
        break;
    } 
  }

  void writeVectorValue(double Vx, double Vy, int mode){}
  void plot(){}
}
\end{verbatim}
\caption{The class {\tt LaplaceGUI.java}, part 4}
\label{fig:PAprogramfour}
\end{center}
\end{figure}
     
The first method in the code is the constructor, the initialization program 
for \t{LaplaceGUI}.  The first line calls the initialization of 
\t{PhysicsApplet},
giving the title, setting the dimensions \t{Nx}, \t{Ny} of the arrays as 
 $100\times 100$, and defining the size of lattice point
as 4 pixels on the screen.  The remaining lines initialize the arrays. 

Most of the remaining methods in the program are abstract methods of the class
\t{PhysicsApplet}.  The task of 
defining these methods guides the programming of the
interface.  The first two of these methods reset one or both of the 
arrays to zero.
The following method, \t{buildPicture}, fills in the picture at 
the center of the
applet.  The method \t{refreshPicture} redraws the picture when this is needed.
(I advise you not to wait for the inscrutable processes of Java to 
decide when to
redraw.)  The method \t{buildControls} sets up the array of buttons at the 
bottom of the applet. \t{PhysicsApplet} defines two types of buttons, a 
\t{CommandButton} that executes a specific command and a \t{ModeButton} that
turns on a specific mode, for example, for drawing on the screen.  Commands 
and other actions requested by clicking the mouse are handled by 
\t{doAction}.  The three methods immediately following \t{buildControls}
implement application-specific parts of the methods for drawing in an
\t{ArrayDisplay}. 

The method \t{WriteVectorValue} is an abstract method of 
\t{PhysicsApplet} that is
not needed by the Laplace Applet.  The method \t{Energy}, which returns the
electrostatic energy, is abstract at this level and is 
to be filled in by the
student.  This method is called in one of the lines of \t{doAction}.

At this point, the class \t{LaplaceGUI} has defined all of the abstract methods
of the class \t{PhysicsApplet} except the crucial physics method \t{solve}.  
Once this 
method and the new method \t{Energy} are defined, the applet can be 
created and brought to the screen by the standard procedures of the \t{Applet}
class.  So the only task left for the
program \t{Laplace.java} is to define these
two methods.

The construction of \t{LaplaceGUI} has some tedious components, but this is 
the irreducible tedium of making sure that all of the buttons and 
controls needed
for the analysis are in place.   
Once this job is done, all that is left to the student
is to actually program the physics.

\section{Conclusions}

In this paper, I have described a system for programming numerical exercises
to accompany core undergradudate physics courses.  A class library 
\t{PhysicsApplet} supplies the underlying graphical and control elements.
Using this resource, an instructor would write a program that defines the 
visual form of the numerical calculation as a Java applet.  The actual
programming of the numerical algorithm is left to the student.  The hierarchial
structure of the object-oriented Java programming language makes this 
system straightforward to implement.  
I hope that this model is one that will be
helpful to many instructors in integrating numerical calculations into their
teaching.

\Acknowledgements
I am grateful to Patricia Burchat, Blas Cabrera, Norman Graf, and 
Tony Johnson for discussions of the subjects presented here, and to the 
students in Physics 120--121--122 at Stanford University, whose help and 
feedback was essential in developing these materials.

\appendix

\section{Documentation of the parent class \t{PhysicsApplet}}

Java applets of the kind discussed in this paper can be created by assembling
graphical elements and controls defined in the parent class 
\t{PhysicsApplet}.
This is an abstract class in which the graphical elements appear as embedded
classes.  These elements are initialized and placed into the applet by defining
the abstract methods of \t{PhysicsApplet}.  In this appendix, 
I document the
various elements that \t{PhysicsApplet} makes available and list the abstract
methods that must be defined in order for a daughter class to be instantiated.

In the accompanying Java code, the file \t{PhysicsApplet.java} is contained in 
the directory \t{Templates}.  This directory also contains a file called 
\t{PhysicsApplet.h}. Unlike \t{C}, Java does not make use of \t{.h} files.
But the user can find in this file a list of the important
 variables and methods
of all of the classes defined by \t{PhysicsApplet}.

\subsection{Global form and parameters}

\t{PhysicsApplet} imposes the general form that an applet should have a 
title in a bar at the top, a picture in the center, and a 
set of controls at the 
bottom.  The picture might typically display a numerical computation on a 
grid.  For the user, the important global variables are:
\begin{tabbing}
 \t{int pixelsizesizesize}  \=    \kill
\t{int Nx} \> grid points in $x$ \\
\t{int Ny} \> grid points in $y$ \\
\t{int pixelsize} \> pixels on the screen/grid point \\
\end{tabbing}
The class \t{PhysicsApplet} has a constructor that supplies the data for these 
variables.  The constructor for a daughter class of \t{PhysicsApplet} should 
call this constructor by having as its first line
\beq
\verb!super("Applet Title",Nx,Ny,pixelsize,extraxsize);!
\eeq{constructor}
as shown in Fig.~\ref{fig:FTprogram}.  The entry \t{extrasize} should equal
the extra white space to be left around the picture, in pixels.

The constructor for the applet then places the title at the top, 
calls a routine
\beq
\verb!buildPicture();!
\eeq{buildPic}
and calls a routine
\beq
\verb!buildControls();!
\eeq{buildCon}
to set up the array of buttons.  These are abstract methods of 
\t{PhysicsApplet}.
A daughter class can define these methods by making use of the graphical 
elements described in the next two sections.

The size of the applet on the screen is actually determined by the information 
in the \t{html} file called by the Web browser.  The file \t{Laplace.html}
that controls the Laplace applet discussed in Section 2 has as its entire 
content:
\beq
\verb!<applet code="Laplace.class" width="500" height="600"></applet>! 
\eeq{htmlfile}
The file specifies the width and height of the applet, in pixels.  
It is usually 
necessary to adjust these values so that the applet appears with the best
size.  The file \t{Laplace.class} is the compiled program from 
\t{Laplace.java}.
The compilations generates many other associated \t{class} files; if these are 
in the same directory as the \t{html} file, the browser will pick
 these up when the 
program is run.  Alternatively, it is possible to collect these 
\t{class} files in a
single `Java archive' or \t{jar} file.  For the problem set exercises 
described here,
I posted on the course Web page, in addition to the basic source code,
 the \t{html} file, the \t{class} file for the original (broken) form 
of the applet, and 
a \t{jar} file containing the remaining \t{class} files needed to implement the
applet.  Then, when a student accesses the \t{html} file on this Web page, the
applet loads and displays its original behavior.

\subsection{Controls}

To make programming the graphical user interface as easy as possible, 
the operation of the applet is modal.  An integer \t{mouseMode} is a 
variable of \t{PhysicsApplet}.  This variable controls the various
mouse actions.  Functions called by the mouse actions, e.g., 
\t{writeFieldValue} defined below, should test for the correct
value of the \t{mouseMode} before performing the action.

The \t{mouseMode} can be changed by a \t{ModeButton}.  The constructor
is
\beq
\verb!ModeButton(String name, int ModeCode)!
\eeq{ModeBt}
The variables specify the name of the button and the value to which
 \t{mouseMode} should be set.

Another purpose of a button is to execute a command. All button commands 
are executed by the abstract method of \t{PhysicsApplet}
\beq
\verb!doAction(Code)!
\eeq{doA}
When \t{doAction} is defined, 
the integer argument should go to a \t{switch} statement, and the command
associated with the given code should then be executed.  A button that 
called \t{doAction} with a given code is constructed by 
\beq
\verb!CommandButton(String name, int CommandCode)!
\eeq{CommB}

It is often useful to include a scrollbar to control some physical 
variable.  The constructor
 \beq
\verb!PhysicsScrollbar(Value, Minimum, Maximum, CommandCode)!
\eeq{Scrbar}
creates a scrollbar with response between the values \t{Minimum} and 
\t{Maximum}.  The default value is \t{Value}; the code to read the 
scrollbar is \t{CommandCode}.  The scrollbar defined by this method
uses a logarithmic scale internally, since this gives better control 
in parameter adjustment.

All three of these items can be directly included in the control panel
\beq
   \verb!Panel Controls!
\eeq{controls}
using the \t{Panel} method \t{add()}.  This would be done in the 
definition of the method \t{buildControls}.

The applet also defines a Java class called a \t{Thread} that controls an
abstract process.  In the \t{PhysicsApplet}, this \t{Thread} executes the 
method \t{solve()}. More specifically, calling
\beq
  \verb!startThread();! 
\eeq{starting}
lauches the \t{solve()} method.  Calling
\beq
  \verb!stopThread();! 
\eeq{stopping}
causes a variable \t{timetostop} to be set to \t{true}.  
If \t{solve()} contains a loop,
one can check for this condition and exit the loop if 
it is satisfied.  This is done,
for example, in the program \t{Laplace.java} shown in Fig.~1.

\subsection{Graphical elements}

The graphical elements supplied by \t{PhysicsApplet} are embedded classes 
of this parent class.  In this section, I describe these elements and 
list their
public methods.

\subsubsection{\t{TitleBanner}}

A \t{TitleBanner} is a component that holds one line of text.  
A \t{TitleBanner}
is initialized by writing
\beq
\verb!myBanner = new TitleBanner("Title",24);! \   ,
\eeq{mytitle}
giving the initial text string and the font size.  To change the 
text string, call
\beq
\verb!myBanner.write("New Text");!\  .
\eeq{newtitle}

\subsubsection{\t{ArrayDisplay}}

An \t{ArrayDisplay} is a component that displays the values of an array in 
grayscale.  The underlying data for this class are an array of doubles and an 
array of integers, both indexed from \t{0} to \t{Nx}, and from \t{0} to \t{Ny}
(\t{(Nx+1)(Ny+1)} components).  An \t{ArrayDisplay} is initialized by writing
\beq
\verb!myAD = new ArrayDisplay(A, ID, Color.red, Color.blue);! \   ,
\eeq{myarray}
where \t{A} is the array of doubles, \t{ID} is the array of integers.  The 
constructor of the \t{ArrayDisplay} creates two vectors of colors, 
\beq
\verb!DisplayColors[i]! \ \mbox{and}\ \verb!altDisplayColor[i]! \ ,
\eeq{colorarrays}
indexed over \t{ i  = 0 ... 10}.

For an \t{ArrayDisplay} \t{AA}, calling \t{AA.refresh()} 
causes the display to be 
redrawn on the screen with the current values of the array.

To operate to \t{ArrayDisplay}, it is necessary to 
define three functions which are
abstract methods of \t{PhysicsApplet}:
\beqa
& &\verb!Color findColor(double a, int id)!\CR
& &\verb!void writeFieldValue(double Val, int mode)!\CR
& &\verb!void setArrays(int i, int j, int mode)!
\eeqa{Arrayfunctions}
The first of these functions takes a value \t{a} of an element of \t{A} 
and the value \t{id} of the corresponding element of \t{ID}
and returns the color
that the corresponding cell should be painted.  The second takes a 
computed value and a mode number and is expected to issue a command to 
write the value to the screen.  The third takes a coordinate pair \t{(i,j)}
and a mode number and is expected to set the corresponding element of
\t{A} or \t{ID} to a fixed value. The three methods are called by the mouse
operations associated with the \t{ArrayDisplay}. 
 All three methods are illustrated in 
the implementation
of \t{LaplaceGUI} described in Section 4.

\subsubsection{\t{VectorDisplay}}

A \t{VectorDisplay} is a component that displays the values of two arrays
as vectors on the screen.   The underlying data for this class are
two arrays of doubles and an 
array of integers, all indexed from \t{0} to \t{Nx}, and from \t{0} to \t{Ny}
(\t{(Nx+1)(Ny+1)} components).  A \t{VectorDisplay} is initialized by writing
\beq
\verb!myVD = new VectorDisplay(Bx, By, ID, bscale, Color.red, Color.blue);! 
\   ,
\eeq{myVarray}
where \t{Bx}, \t{By} are the two arrays of doubles, \t{ID} is the array of 
integers, and \t{bscale} is an typical scale for the length of the
vectors.

For a \t{VectorDisplay} \t{VV}, calling \t{VV.refresh()} 
causes the display to be 
redrawn on the screen with the current values of the arrays.  Calling
\t{VV.resetscale(nbs)} causes the display to be redrawn with the reference 
vector length \t{bscale} set to \t{nbs}.

To operate a \t{VectorDisplay}, it is necessary to 
define two functions that are
abstract methods of \t{PhysicsApplet}:
\beqa
& &\verb!void writeVectorValue(double Vx, double Vy, int mode)! \CR
& &\verb!void setArrays(int i, int j, int mode)!
\eeqa{Vectorfunctions}
The first of these functions takes computed vector components
and a mode number and is expected to issue a command to 
write the value to the screen.  The second takes a coordinate pair \t{(i,j)}
and a mode number and is expected to set the corresponding element of
\t{Bx}, \t{By}, or \t{ID} 
 to a fixed value.   The three methods are called by the mouse
operations associated with the \t{VectorDisplay}. Their operation is very
similar to that for the corresponding functions in \t{ArrayDisplay} just
above.

\subsubsection{\t{CurveDisplay}}

A \t{CurveDisplay} is a component that displays the values of two 
functions of $x$.  The underlying data for this class are two vectors \t{f[n]}
and \t{h[n]}, both indexed from \t{0} to \t{Nx}.  A \t{CurveDisplay} is 
initialized
by writing
\beqa
& &\verb!myCD = new CurveDisplay(Height,Zero,f,h,fscale,hscale, !\CR
 & & \hskip 3.0in \verb!fcolor, hcolor, Mode);!
\eeqa{CDdef}
where \t{Height} is the height of the display in pixels, 
\t{Zero} is the vertical 
position of $f = 0$ in pixels, \t{f} and \t{h} are the basic 
data vectors, \t{fscale}
and \t{hscale} are anticipated maximum values of the functions, \t{fcolor} and 
\t{hcolor} are Java Color classes for each function  ({\eg}, 
\t{Color.red}), and 
\t{Mode} is an integer.  When the global variable \t{MouseMode} is set to this
value, the user can redraw the function \t{f} with the mouse.

For a \t{CurveDisplay} \t{CC}, calling \t{CC.refresh()} causes the 
display to be 
redrawn on the screen with the current values of the arrays.  Calling
\t{CC.resetfscale(newfscale)} or  \t{CC.resethscale(newhscale)}
causes the display to be redrawn with the a change in the scale of the 
corresponding function.

\subsubsection{\t{PlotDisplay}}

A \t{PlotDisplay} is a component that displays a plot.   A \t{PlotDisplay} is
initialized by writing
\beq
\verb!myPD = new PlotDisplay(Xa,Xb,Ya,Yb,Xtick,Ytick);!
\eeq{PDdef}
where the limits of the plot are specified as \t{Xa} to \t{Xb}, 
\t{Ya} to \t{Yb}, and
the ticks on the edge of the plot in x and y are spaced by \t{Xtick}, 
\t{Ytick}.

For a \t{PlotDisplay} \t{PP}, calling \t{PP.refresh()} causes the display 
to be 
redrawn.

The plot is actually drawn by the method 
\beq
  \verb!void plot()!
\eeq{toplot}
which is an abstract method of \t{PhysicsApplet}.  Points, text, and lines are
drawn into the plot in \t{PP} by including the following calls in the 
body of \t{plot}:
\beqa
& & \verb!PP.verticalAxis(x);!\CR
& & \verb!PP.horizontalAxis(y);!\CR
& & \verb!PP.plotPoint(x,xerror,y,yerror)!\CR
& & \verb!PP.drawLine(x1,y1,x2,y2)!\CR
& & \verb!PP.drawString("The Text",x,y);!
\eeqa{basicplot}

To plot a function, the \t{PlotDisplay}  makes use of a helper class called a
\t{plot\-Stream}.  This is, essentially, a collection of points.
 To create a \t{plotStream} inside a \t{PhysicsApplet}, simply call
\beq
  \verb!plotStream PS = new plotStream()!;
\eeq{PSbegin}
Points are added to a \t{plotStream} by the commands
\beqa
& & \verb!PS.add(x,y);!\CR
& & \verb!PS.add(x, y, yerror)!\CR
& & \verb!PS.add(x,xerror,y,yerror)!
\eeqa{addtoPS}
The number of elements in a \t{plotStream} is returned
 by \t{PS.size()}.  However,
the \t{plotStream} is designed so that the whole set of points can be plotted
with one command.  This can be done in a number of formats, by the 
\t{PlotDisplay} commands:
\beqa
& & \verb!PP.plotPoints(PS);!\CR
& & \verb!PP.plotLines(PS);!\CR
& & \verb!PP.plotCurve(PS);!\CR
& & \verb!PP.plotHistogram(PS);!
\eeqa{PSmethods}
The first of these plots the points as data points, 
with error bars if the errors
have been provided.  The second plots the points as a broken-line function.
The third plots the points by forming a smooth curve with a cubic spline.
The fourth plots the points as a histogram, interpreting the x 
coordinates as the
bin centers.

To change the plotting color to red, call 
\t{PP.setColor(Color.red)}; any \t{Color}
defined by Java may be used in the same way.  The call \t{PP.switchColor()}
cycles through the possible colors.

\subsection{Abstract methods of \t{PhysicsApplet}}

To recapitulate, I list the abstract methods of \t{PhysicsApplet} that
must be defined in a daughter class.  If a method is not needed for the
particular applet being constructed, it should be defined in a trivial
way, {\eg}, with a body that is null or contains only \t{return 0}.
\beqa
& & \verb!abstract void resetArrays();!\CR
& & \verb!abstract void resetAll();!\CR
& & \verb!abstract void buildPicture();!\CR
& & \verb!abstract void refreshPicture();!\CR
& & \verb!abstract void buildControls();!\CR
& & \verb!abstract void doAction(int Code);!\CR
& & \verb!abstract void solve();!\CR
& & \verb!abstract Color findColor(double A, int S);!\CR
& & \verb!abstract void writeFieldValue(double Val, int mode);!\CR
& & \verb!abstract void writeVectorValue(double Vx, double Vy, int mode);!\CR
& & \verb!abstract void setArrays(int i, int j, int mode);!\CR
& & \verb!abstract void plot();!
\eeqa{abstrr}

\section{Guide to the accompanying Java code}

The Java code of \t{PhysicsApplet.java} and  some example applets 
can be found as a \t{tar} file submitted with the eprint of this 
paper~\cite{eprint}.  This \t{tar} file unpacks to a set of ten 
 directories.  One of these,
\t{Templates}, contains the file \t{PhysicsApplet.java}.  The other nine 
each contain an example applet. 

Each of these applets is given in the following form:  If the name of the 
directory is \t{B}, the directory will contain files \t{B.java}, \t{BGUI.java},
\t{B.html}, \t{myB.java}, and \t{myB.html}.  \t{PhysicsApplet.java} must be 
copied into the directory before compiling.  The program \t{B.java} is the 
program given to students to complete.  To compile it (on a UNIX system)
type \t{javac B.java}; to run it, type \t{appletviewer B.html}.  The program
\t{myB.java} is the completed, working applet.  It can be compiled and run
in the same way.

The nine applets included in the distribution are the following:
\begin{itemize}
\item \t{Laplace}:  An applet that solves the 
 Laplace equation in two dimensions.  This applet was described in Section 2.
\item \t{Dielectric}: An applet that solves electrostatic problems with 
dielectric
   material in two dimensions
\item \t{Magnet}: An applet that solves magnetostatic problems in two 
dimensions.  This applet was described in Section 3.
\item \t{FourierLab}: An applet that illustrates the Fourier Transform.  This
applet was described in Section 3.   This directory includes an implementation
of the Fast Fourier Transform adapted to the applet. 
To use it, change the name of the file
\t{myFastFourierTransform.java} to \t{myFourierTransform.java}.
\item \t{Wave}: An applet that solves a simplified form of the
one-dimensional wave equation 
$(\partial/\partial t - \partial/\partial x) \phi = 0$ (following
\cite{Numerical}).
\item \t{Disperse}: An applet that solves the one-dimensional wave equation 
by using the Fourier Transform, with an arbitrary input dispersion relation.
\item \t{Bessel}: An applet that  illustrates the numerical computation of 
the Bessel functions $J_0(z)$ and $J_1(z)$.
\item \t{Antennae}: An applet that computes the radiation pattern from an 
array of antennae.
\item \t{Diffraction}: An applet that, given an aperture drawn on the screen,
computes its Fraunhofer diffraction pattern.
\end{itemize}

\end{document}